\documentclass[twocolumn,showpacs,superscriptaddress,amsmath,amssymb]{revtex4}
\usepackage{graphicx}
\usepackage{dcolumn}
\usepackage{bm}

\usepackage{amsmath}
\usepackage{amssymb}

\newcommand{\UA}{\mbox{$U_{A}(1)$}}
\newcommand{\qbar}{\mbox{$\bar q$}}
\newcommand{\ubar}{\mbox{$\bar u$}}
\newcommand{\dbar}{\mbox{$\bar d$}}
\newcommand{\sbar}{\mbox{$\bar s$}}

\newcommand{\bea}{\begin{eqnarray}}
\newcommand{\eea}{\end{eqnarray}}
\newcommand{\la}{\left\langle}
\newcommand{\ra}{\right\rangle}
\begin{document}

\title{$\eta$- \& $\eta'$-mesic nuclei and $U_A(1)$ anomaly at finite density}

\author{Hideko Nagahiro}
   \affiliation{Research Center for Nuclear Physics(RCNP), Osaka University, 
Ibaraki, Osaka, 567-0047, Japan
}
\author{Makoto Takizawa}
   \affiliation{Showa Pharmaceutical University, Machida, Tokyo,
 194-8543, Japan}

\author{Satoru Hirenzaki}
   \affiliation{Department of Physics, Nara Women's University, Nara, 630-8506, Japan}

\begin{abstract}
We discuss theoretically the possibility of observing the bound states of
 the $\eta$ and $\eta'(958)$ mesons in nuclei.
We apply the NJL model to study the $\eta$ and $\eta'$ meson properties
 at finite density
and calculate the formation cross sections
 of the $\eta$ and $\eta'$ bound states with the Green function method
 for ($\gamma$,p) 
 reaction. We also discuss the experimental feasibility at photon facilities
 like SPring-8.
The contributions due to the $\omega$ meson production are also included
 to obtain the realistic ($\gamma$,p) spectra.
We conclude that 
we can expect to observe resonance peaks in ($\gamma$,p) spectra for the
 formation of meson bound states and we can deduce new
 information on $\eta$ and $\eta'$ properties at finite density. These
 observations are believed to be essential to know the possible mass
 shift of 
 $\eta'$ and deduce new information on the effective restoration of the
$U_A(1)$ anomaly in the nuclear medium.
\end{abstract}


\maketitle

\section{Introduction}
\label{}


Understanding the low-lying hadron structures from the viewpoint of
the quark and gluon degrees of freedom is one of the most challenging 
problems in the quantum chromodynamics (QCD) because of its
nonperturbative nature in the low-energy regime.

The lightest excitation on the QCD vacuum is the pion which is considered as 
a quark and an antiquark bound state in the pseudoscalar channel. 
Its mass is off-scale light compared with other hadrons. 
This can be understood by recognizing that the chiral symmetry is 
spontaneously broken in the QCD vacuum and the pion is the 
Nambu-Goldstone (NG) boson associated with the dynamical chiral 
symmetry breaking (DCSB). It is believed to be responsible for a large part of 
the constituent quark masses, which are introduced in the many constituent 
quark models.
The effects of the explicit breaking of the chiral symmetry
on the pion properties have been systematically studied using the effective 
lagrangian composed of the pion field. This approach is called the chiral
perturbation theory (ChPT) \cite{GL1984}. The success of ChPT approach 
supports the importance of the DCSB in low-energy QCD.

When we start looking at the strange quark sector, 
we encounter another problem. 
The ninth heavier pseudoscalar meson is $\eta'$ and its mass is 
much heavier than other octet pseudoscalar mesons. 
Weinberg showed that the mass of $\eta'$ should be less
than $\sqrt{3} m_{\pi}$ if \UA\ symmetry were not
explicitly broken \cite{Weinberg1975}.
Thus the \UA\ symmetry must be broken.
The key step to solve this \UA\ problem was to realize that 
\UA\ symmetry was explicitly broken by the quantum anomaly.
In the following year, 't Hooft pointed out the relation between 
\UA\ anomaly and topological gluon configurations of QCD, i.e., 
instantons and showed that the interaction of light quarks and 
instantons breaks the \UA\ symmetry \cite{tHooft1976}.
He also showed that such an interaction can be represented by a
local $2N_{f}$ quark vertex, which is antisymmetric under
flavor exchanges, in the dilute instanton gas approximation.

The effects of the \UA\ anomaly on the low-energy QCD have been
extensively studied in the $1/N_c$ expansion approach \cite{Nc}.
In the $N_c \rightarrow \infty$ limit, the \UA\ anomaly is turned off
and then the $\eta$ and $\eta'$ mesons become the ideal mixing states.
The higher order effects of the $1/N_c$ expansion
give rise to the flavor mixing between the $\eta$ and $\eta'$ mesons 
and push up the $\eta'$ mass  \cite{RST80,DiVe80,KO80,NA81}. 
They were further discussed in the 
context of the ChPT \cite{NcChPT} and the reasonable description of 
the nonet pseudoscalar mesons was obtained.

The \UA\ breaking $2 N_f$ quark determinant interaction was 
introduced to the low-energy effective quark models of QCD.
The low-lying meson properties have been studied 
\cite{KH1988,BJM1988,KKT1988,RA1988,KLVW1990,TO1995,NOT1996} using 
the three-flavor version of the Nambu-Jona-Lasinio (NJL) model \cite{NJL}
with the Kobayashi-Maskawa-'t~Hooft 
(KMT) determinant interaction \cite{tHooft1976,KM1970}.  
It is further argued that the \UA-breaking interaction gives
rise to the spin-spin forces, which are important for light baryons 
\cite{SR1989,OT1989,MT1993}.

It was observed by Witten that the instanton contribution
scales as $\exp(-N_c)$ which seems to contradict the
$1/N_c$ expansion approach, $m^2_{\eta'} = O(1/N_c)$~\cite{NPB149}.
A few years ago, Sch\"afer showed that the large $N_c$ behavior of
the instanton approach is consistent with the $1/N_c$ expansion approach
if the instanton ensemble is stabilized by a classical repulsive
core~\cite{PRD66}. 
The use of the $U_A(1)$ breaking quark determinant interaction with the
low-energy effective quark model of QCD in the mean field approximation
seems to mimic the dilute instanton liquid picture, and therefore, we
consider
this approach is not inconsistent with the $1/N_c$ expansion approach.

The dynamics of instantons in the multi-instanton vacuum has
been studied by many authors, either in the models or in the
lattice QCD approach, and the widely accepted picture is that
the QCD vacuum consists of small instantons of the size about
1/3 fm with the density of 1 instanton (or anti-instanton) per
fm$^{4}$ \cite{Instanton}. In the instanton liquid model, 
the instanton plays a crucial role in understanding not only the 
\UA\ anomaly but also the spontaneous breaking of chiral symmetry itself.

In this paper, we consider the $\eta$ and $\eta'$ meson bound states in
finite nuclei, {\it i.e.}, $\eta$- and $\eta'$-mesic nuclei in order to
investigate the change of the QCD vacuum structure at finite density
through the $\eta$ and $\eta'$ meson properties in the nuclear
medium. Since the vacuum properties of $\eta$ and $\eta'$ are believed
to have close relation to the QCD vacuum structure as described above,
we can expect to have new information and much deeper insights on the
QCD vacuum by knowing the $\eta$ and $\eta'$ meson properties at finite
density, where the partial restoration of the DCSB is expected.

The $\eta$-mesic nucleus was studied by Haider and Liu~\cite{PLB172etc}
and Chiang, Oset and Liu~\cite{PRC44_eta}. As for the formation
reaction, the attempt to find the bound states by ($\pi^+$,p) reaction
led to a negative result~\cite{PRL60}. Recently, the $\eta$ meson and
light nuclei systems are studied experimentally~\cite{PRL92_eta3He} and
also theoretically~\cite{Jain}. The experimental study in
Ref.~\cite{PRL92_eta3He} claims the existence of the $\eta$-bound state
in the light nucleus.

Recently, 
there are several very important developments
in the research of the
spontaneous breaking of chiral symmetry and its partial restoration at
finite density
by studying
the hadronic systems, such as pionic atoms~\cite{PLB514etc,PRL92,APPB31etc}, 
$\eta$- and $\eta'$-mesic
nuclei~\cite{PLB443,EPJA6,our_eta,valencia_eta,PRL94,PLB634} and $\omega$-mesic 
nuclei~\cite{PLB443,EPJA6,PRC59,NPA650,NPA706}
in both of theoretical and experimental aspects.
There is also a theoretical study on $\sigma$ meson  
nuclear bound states
based on SU(2) linear sigma model~\cite{NPA710}.
Especially,
after a series of deeply bound pionic atom experiments~\cite{PRC62etc,PRL88},
K.~Suzuki {\it et al.} reported the quantitative
determination of pion decay constant $f_\pi$ in-medium~\cite{PRL92}  from the deeply
bound pionic states in Sn isotopes~\cite{PTP103} and stimulated many active researches
of the partial restoration of chiral symmetry at finite
density~\cite{PLB514etc,APPB31etc,PLB541,PLB563,PLB578}.

However, as for the behavior of the $U_A(1)$ anomaly in the nuclear
medium, 
the present exploratory level is rather poor. Although some theoretical
results have been reported, there exists no experimental information on
the possible effective restoration of the $U_A(1)$ anomaly at finite
density.
In the context of the instanton dynamics, the mass of
the $\eta'$ at a nonzero baryon chemical potential was
studied in the two color QCD~\cite{PRD67}.
As for the case of the finite temperature, the temperature
dependence of the instanton density was calculated
in~\cite{RMP53,PRD29,PLB218,PRD53,PRD50}.
In the low-energy effective model of QCD approach, 
T.~Kunihiro studied the effects of the $U_A(1)$ anomaly on
$\eta'$ properties at finite temperature using the
NJL model with the KMT term and showed the possible character
changes of $\eta'$ at $T\ne 0$ \cite{PLB219}.
Theoretical predictions by
other authors also
supported the possible change of the $\eta'$ properties at
finite density as well as at finite temperature~\cite{PRC63,PLB560etc}.
Experimental feasibilities to observe the $\eta'$-mesic nuclei were also
studied in this context
in our previous work in Ref.~\cite{PRL94}, where the in-medium 
$\eta'$ properties were estimated quite roughly based on the results
reported in Refs.~\cite{PLB219,PRC63,PLB560etc}.
In Ref.~\cite{PLB634} further theoretial studies have been performed for
$\eta$- and $\eta'$-nucleus systems with paying attention to the QCD
axial $U(1)$ problem.

In this study,
we consider the $\eta$ and $\eta'$ meson properties in atomic nuclei and
the structures of the $\eta$- and $\eta'$-mesic nuclei using the
SU(3)$_f$ NJL model
in order to get much deeper insights
on the $\eta'$ mass shifts and their relation to  $U_A(1)$  
anomaly
effects
and to get more quantitative results than  our previous work~\cite{PRL94}.
We also
propose the formation reaction of the $\eta$- and $\eta'$-mesic
nuclei and discuss the possibility to produce the $\eta$- and
$\eta'$-nucleus bound states.
The $\eta$ and $\eta'$ properties
in the medium should provide
us important information on the
QCD vacuum structure, especially on the
effective restoration of the
$U_A(1)$ symmetry in the nuclear medium.

This paper is organized as follows.
In Sec.\ref{form}, we will obtain the quark masses and the quark
condensates in 
finite density using NJL model, and calculate the meson masses
in-medium. In Sec.\ref{spectra}, we will discuss the formations of mesic
nuclei using the Green function method, and show the numerical results
of the missing mass spectra of the ($\gamma$,p) reaction. Finally,
Sec.\ref{conc} is devoted to summary of this paper.

\section{Quark and meson masses in NJL model}
\label{form}%
We work with the following NJL model lagrangian density extended to 
three-flavor case:
\begin{subequations}
\begin{eqnarray}
\mathcal{L} & =  &\mathcal{L}_0 + \mathcal{L}_4 + \mathcal{L}_6 , \label{njl1} \\
\mathcal{L}_0 & = & \bar \psi \,\left( i \partial_\mu \gamma^\mu - \hat m 
\right) \, \psi \, ,
\label{njl2} \\
\mathcal{L}_4 & = & {\frac{g_S}{2}} \sum_{a=0}^8 \, \left[\, \left( 
\bar \psi \lambda^a \psi \right)^2 + \left( \bar \psi \lambda^a i \gamma_5 
\psi \right)^2 \, \right] \, ,
\label{njl3} \\
\mathcal{L}_6 & = & g_D \left\{  {\rm det} \left[ \bar \psi_i (1 - \gamma_5) 
\psi_j \right] + 
h.c.
\right\} \, .
\label{njl4}
\end{eqnarray}
\end{subequations}
Here the quark field $\psi$ is a column vector in color, flavor and Dirac 
spaces and $\lambda^a (a=0\ldots 8)$ is the Gell-Mann matrices for 
the flavor $U(3)$. 
The free Dirac lagrangian $\mathcal{L}_0$ incorporates the current quark mass 
matrix $\hat m = {\rm diag}(m_u, m_d, m_s)$ which breaks the chiral 
$U_L(3) \times U_R(3)$ invariance explicitly. $\mathcal{L}_4$ is a QCD 
motivated four-fermion interaction, which is chiral $U_L(3) \times U_R(3)$
invariant.  The Kobayashi-Maskawa-'t Hooft determinant interaction
$\mathcal{L}_6$  
represents the $U_A(1)$ anomaly.  
It is a $3 \times 3$ determinant with respect to flavor with 
$i,j = {\rm u,d,s}$.   

Quark condensates $\la \qbar q \ra$ and constituent quark masses 
$M_q$ are self-consistently determined by the gap equations 
in the mean field approximation,
\bea
M_u & = & m_u - 2g_S \la \ubar u \ra - 
  2 g_D \la \dbar d \ra \la \sbar s \ra \, , \nonumber \\
M_d & = & m_d - 2g_S \la \dbar d \ra - 
  2 g_D \la \sbar s \ra \la \ubar u \ra \, , \nonumber \\
M_s & = & m_s - 2g_S \la \sbar s \ra - 
  2 g_D \la \ubar u \ra \la \dbar d \ra \, , \label{gap}
\eea
with 
\begin{eqnarray}
\la \qbar q \ra & = & - {\rm Tr}^{(c,D)} \left[ iS_F^q (x = 0) \right] 
\nonumber \\
& = & - \int \frac{d^4p}{(2\pi)^4} {\rm Tr}^{(c,D)}
\left[ \frac{i}{p_\mu \gamma^\mu - M_q + i\epsilon} \right] 
\nonumber \\
& = & - 2 N_c \int^\Lambda\frac{d^3p}{(2\pi)^3}\nonumber\\
&&\times
\frac{M_q}{E_q}
\left( 1 - n_p(T, \mu_q) - \bar n_p(T, \mu_q) \right) 
\, .
\label{condensate}
\end{eqnarray}
Here, the ultraviolet cutoff $\Lambda$ is introduced to regularize the
divergent integral and Tr$^{(c,D)}$ means trace in color and Dirac spaces.
$E_q = \sqrt{ \boldsymbol{p}^2_q + M_q}$ is the on-shell energy of the quark and 
$n_p$ and $\bar n_p$ are Fermi occupation numbers of quarks and 
antiquarks, respectively, defined as,
\bea
n_p(T, \mu_q) & = & \frac{1}{1+{\rm e}^{(E_q - \mu_q)/T}} \, , \nonumber \\
\bar n_p(T, \mu_q) & = & \frac{1}{1+{\rm e}^{(E_q + \mu_q)/T}} \, ,
\label{eq:qnd}
\eea 
where
$T$ represents the temperature of the system and $\mu_q$ is the 
quark chemical potential.
$n_p$ and $\bar{n}_p$ depend on the momentum $\boldsymbol{p}_q$ through
the energy $E_q$.
For $T=0$, we can simply write Eqs.~(\ref{eq:qnd}) as 
\bea
n_p(0, \mu_q) & = & \theta(\mu_q-E_q),\nonumber \\
\bar n_p(0, \mu_q) & = & 0,
\eea 
and by integrating them in momentum space we obtain the quark number
density $\rho_q$ as 
\begin{equation}
\rho_q=\frac{1}{\pi^2}(\mu_q^2-M_q^2)^{3/2}\theta(\mu_q-M_q).
\end{equation}
To simulate the symmetric nuclear matter,
we consider SU(2) symmetric quark matter defined as $\rho_u=\rho_d$,
$\rho_s=0$.
In this case, the nucleon density $\rho$ is defined as,
\begin{eqnarray}
\rho&=&\frac{1}{3}(\rho_u+\rho_d)\nonumber\\
&=&\frac{2}{\pi^2}(\mu_u^2-M_u^2)^{3/2}\theta(\mu_u-M_u),
\label{eq:rho}
\end{eqnarray}
where $\mu_u=\mu_d$ and $M_u=M_d$.
The NJL model at finite temperature and density has been
reviewed in \cite{NJLREV}.

The pseudoscalar channel quark-antiquark scattering amplitudes,
\begin{equation}
\la p_3 , \bar p_4 ; {\rm out} \right. \left| p_1 , \bar p_2 ; {\rm in} \ra 
 =  (2 \pi)^4 \delta^4(p_3 + p_4 - p_1 - p_2) \mathcal{T}_{q \bar q} 
\end{equation}
are then calculated in the ladder approximation. 
We assume that $m_u = m_d$ so that the isospin symmetry is exact.
In the $\eta$ and $\eta'$ channels, 
the explicit expression of $\mathcal{T}_{q \bar q}$ is 
\begin{eqnarray}
\mathcal{T}_{q \bar q} = &-&
\left(
\begin{array}{c} 
\bar u(p_3) \lambda^8 v(p_4) \\
\bar u(p_3) \lambda^0 v(p_4) 
\end{array} 
\right)^T \,
\left(
\begin{array}{cc}
A(q^2) & B(q^2) \\
B(q^2) & C(q^2) \\
\end{array}
\right) \nonumber \\
&\times&
\left(
\begin{array}{c}
\bar v(p_2) \lambda^8 u(p_1) \\
\bar v(p_2) \lambda^0 u(p_1)
\end{array}
\right) \, , \label{qas1}
\end{eqnarray}
with 
\begin{subequations}
\begin{eqnarray}
A(q^2) &=&\frac{2}{{\rm det}{\bf D}(q^2)} 
\left\{ 2\,\tilde{G}\,I^0 (q^2) - G_8 \right\} ,
\label{qas2} \\
B(q^2)&=& \frac{2}{{\rm det}{\bf D}(q^2)}
\left\{- 2\,\tilde{G}\,I^m (q^2) - G_m \right\} , \label{qas3} \\
C(q^2) &=&\frac{2}{{\rm det}{\bf D}(q^2)}
\left\{ 2 \,\tilde{G}\,
I^8 (q^2) - G_0 \right\} , \label{qas4} 
\end{eqnarray}
\end{subequations}
where $q=p_1+p_2$ and $\tilde{G}=G_0 G_8 - G_m G_m$,
and 
\begin{subequations}
\bea
G_0 & = &\frac{1}{2} g_S - \frac{1}{3} ( 2 \langle \bar uu \rangle + 
\langle \bar ss \rangle ) g_D, \\
G_8 & = & \frac{1}{2} g_S - \frac{1}{6} ( \langle \bar ss \rangle - 4  
\langle \bar uu \rangle ) g_D \, , \\
G_m & = & - \frac{1}{3 \sqrt{2}} ( \langle \bar ss \rangle - 
\langle \bar uu \rangle ) g_D \, .
\eea
\end{subequations}

The quark-antiquark bubble integrals are defined by
\begin{subequations}
\begin{multline}
I^0(q^2) =
  i \int \frac{d^4p}{(2 \pi)^4} \\
\times{\rm Tr}^{(c,f,D)}
\left[ S_F(p) \lambda^0 i \gamma_5 S_F(p+q) \lambda^0  i \gamma_5\right]
\, , \label{int1}
\end{multline}
\begin{multline}
I^8(q^2)  =
 i \int\frac{d^4p}{(2 \pi)^4} \\
\times{\rm Tr}^{(c,f,D)}
\left[ S_F(p) \lambda^8  i \gamma_5 S_F(p+q) \lambda^8  i \gamma_5 \right]
\, , \label{int2} 
\end{multline}
\begin{multline}
I^m(q^2)  =
 i \int\frac{d^4p}{(2 \pi)^4} \\
\times{\rm Tr}^{(c,f,D)}
\left[ S_F(p) \lambda^0  i \gamma_5 S_F(p+q) \lambda^8  i \gamma_5 \right]
\, . \label{int3} 
\end{multline}
\end{subequations}
In the calculations of the quark loop integrals, we have applied 
standard techniques of thermal field theory, i.e., the imaginary-time formalism
and we have introduced the same three-momentum cutoff 
used in the gap equations (\ref{gap}) and (\ref{condensate}).
 
The $2 \times 2$ matrix ${\bf D}$ is given by 
\begin{equation}
{\bf D}(q^2) = 
\left( 
\begin{array}{cc}
D_{11}(q^2) & D_{12}(q^2) \\
D_{21}(q^2) & D_{22}(q^2) 
\end{array}
\right) \, , \label{mat}
\end{equation}
with
\begin{subequations}
\begin{eqnarray}
D_{11}(q^2) & = & 2 G_0 I^0(q^2) + 2 G_m I^m(q^2) - 1 \, , \label{mat11}\\
D_{12}(q^2) & = & 2 G_0 I^m(q^2) + 2 G_m I^8(q^2) \label{mat12} \\
D_{21}(q^2) & = & 2 G_8 I^m(q^2) + 2 G_m I^0(q^2) \label{mat21} \\
D_{22}(q^2) & = & 2 G_8 I^8(q^2) + 2 G_m I^m(q^2) - 1 \, . \label{mat22}
\end{eqnarray}
\end{subequations}
From the pole positions of the scattering amplitude Eq. (\ref{qas1}), the 
$\eta$-meson mass $m_{\eta}$ and the $\eta'$-meson mass $m_{\eta'}$ 
are determined.

    The scattering amplitude Eq. (\ref{qas1}) can be diagonalized by the rotation
in the flavor space 
\begin{eqnarray}
\mathcal{T}_{q \bar q} & = & -
\left(
\begin{array}{c} 
\bar u(p_3) \lambda^8 v(p_4) \\
\bar u(p_3) \lambda^0 v(p_4) 
\end{array} 
\right)^T   {\bf T}_{\theta}^{-1} {\bf T}_{\theta} 
\left(
\begin{array}{cc}
A(q^2) & B(q^2) \\
B(q^2) & C(q^2) \\
\end{array}
\right) {\bf T}^{-1}_{\theta}  \nonumber \\
&& \times {\bf T}_{\theta}  
\left(
\begin{array}{c}
\bar v(p_2) \lambda^8 u(p_1) \\
\bar v(p_2) \lambda^0 u(p_1)
\end{array}
\right) \, , \label{qasm1} \\
& = & -
\left(
\begin{array}{c} 
\bar u(p_3) \lambda^{\eta} v(p_4) \\
\bar u(p_3) \lambda^{\eta'}  v(p_4) 
\end{array} 
\right)^T \, 
\left(
\begin{array}{cc}
D^{\eta}(q^2) & 0 \\
0 & D^{\eta'}(q^2) 
\end{array}
\right) \nonumber \\
&& \times
\left(
\begin{array}{c}
\bar v(p_2) \lambda^{\eta} u(p_1) \\
\bar v(p_2) \lambda^{\eta'} u(p_1)
\end{array}
\right) \, , \label{qasm2}
\end{eqnarray}
with $\lambda^{\eta} \equiv \cos \theta  \lambda^8 - \sin \theta \lambda^0$,
$\lambda^{\eta'} \equiv \sin \theta  \lambda^8 + \cos \theta \lambda^0$ and
\begin{equation}
{\bf T}_{\theta} = \left(
\begin{array}{cc}
\cos \theta & -\sin \theta \\
\sin \theta & \cos \theta 
\end{array}
\right) \, .
\end{equation}
The rotation angle $\theta$ is determined by 
\begin{equation}
\tan 2 \theta = \frac{2 B(q^2)}{C(q^2) - A(q^2)} \, . \label{angle}
\end{equation}
Note that $\theta$ therefore depends on $q^2$.  At $q^2 = m_{\eta}^2$,  
$\theta$ represents the 
mixing angle of the $\lambda^8$ and $\lambda^0$ components in the 
$\eta$-meson state. In the usual effective pseudoscalar meson lagrangian 
approaches, the $\eta$ and $\eta'$ mesons are analyzed using the 
$q^2$-independent $\eta - \eta'$ mixing angle.  Because of the 
$q^2$-dependence, $\theta$ cannot be interpreted as the 
$\eta - \eta'$ mixing angle.  The origin of the $q^2$-dependence 
is that the $\eta$ and $\eta'$ mesons have the internal structures.
%
%

We obtain the dynamical quark and meson masses in vacuum as
compiled in Table ~\ref{tab:parameters} with the input parameters
determined in Ref.~\cite{PRC53}. 
In Ref.~\cite{PRC53}, the 4 model parameters,
namely, cutoff $\Lambda$, 4-quark coupling constant $g_S$, 6-quark coupling 
constant $g_D$ and the current s-quark mass $m_s$ have been fixed so as to 
reproduce the observed values of the pion, kaon, $\eta'$ masses and the pion decay
constant, while the current u, d-quark mass has been fixed at 5.5 MeV, which has been 
taken from the results of the chiral perturbation theory and QCD sum rule approaches.
This parameter set gives an $\eta$ mass of $m_\eta = 514.8$ MeV, which is about 6\%
smaller than the observed value $m_\eta = 547.75$ MeV.  One is able to fit the mass
of $\eta$ instead of the $\eta'$.  The reason why we have used the present parameter
set is just we are more interested in the $\eta'$-mesic nucleus than the $\eta$-mesic
nucleus in this article because we consider that the former is more suitable for 
observing the finite density effect of the $\UA$ anomaly. 

\begin{table}
\begin{tabular}[t]{l|l|l}
\hline
\hline
Input parameters & 
\multicolumn{2}{c}
{Calculated results [MeV]}
\\
\hline
$\Lambda=602.3$ [MeV] & $M_{u,d}=367.6$  & $f_\pi=92.4$  \\
$g_S\Lambda^2=3.67$ & $M_{s}=549.5$  & $m_\pi=135.0$  \\
$g_D\Lambda^5=-12.36$ & $\langle\bar{u}u\rangle^{1/3}=-241.9$  &
 $m_\eta=514.8$  \\
$m_{u,d}=5.5$ [MeV] & $\langle\bar{s}s\rangle^{1/3}=-257.7$  &
 $m_{\eta'}=957.7$  \\
$m_s=140.7$ [MeV] & &\\
\hline
\hline
\end{tabular}
\caption{Input parameters determined in Ref.~\cite{PRC53} and calculated
results in vacuum
in three-flavor NJL model.}
\label{tab:parameters}
\end{table}
In this paper, we investigate the finite density effects on the meson
mass spectra for the following 3 cases with the different strengths
$g_D$ of
the determinant KMT interaction as,
\begin{equation}
\begin{split}
 {\rm (a)}\  & g_D(\rho)=g_D \\
 {\rm (b)}\  & g_D(\rho)=0 \\
 {\rm (c)}\  & g_D(\rho)=g_D \exp[-(\rho/\rho_0)^2],
\label{eq:gD}
\end{split}
\end{equation}
where $g_D$ is the vacuum strength of the determinant interaction as
shown in Table~\ref{tab:parameters}. The 
$g_D(\rho)$ has no density dependence for cases (a) and (b). In case
(a), the meson vacuum properties are well reproduced as shown in Table
\ref{tab:parameters}, while there are no anomaly effects in case (b).
For $g_D=0$ case, we use slightly different parameter set as shown in
Table~I in Ref.~\cite{PRD71} to reproduce the meson masses and the pion
decay constant in vacuum without anomaly effect.
In case (c), we simply assume the density dependence of $g_D$ as this
form in 
order 
to examine the medium effect due to 
density dependence of $g_D$ itself on the meson mass spectra in finite
density. 

Here it may be interesting for our study to notice that
there are theoretical suggestions about 
possible density dependence of $g_D$~\cite{NPA642etc,PTP114}. 
In Ref.~\cite{PTP114}, the effective coupling constant of the
instanton-induced interaction is suggested to have chemical potential
dependence for $N_f=2$ systems. For $N_f=3$ systems, we can expect to
have the similar $\mu$ dependence, though it is not easy to show
explicitly.
We are interested in studying the effect of such density dependence 
discussed in Ref.~\cite{PTP114} on 
meson mass spectra as future works.

\begin{figure}
\includegraphics[width=8cm]{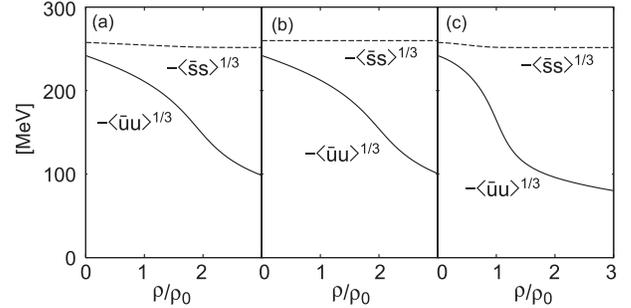}
\caption{Density dependence of the quark condensates
in the SU(2) symmetric matter, $\rho_u=\rho_d$ and $\rho_s=0$, where
$\langle \bar{u}u \rangle = \langle \bar{d}d \rangle$. Three panels
correspond to the cases (a), (b) and (c) defined in Eq.~(\ref{eq:gD}),
respectively. 
The nucleon density $\rho$ is
defined in Eq.~(\ref{eq:rho}) and $\rho_0$ is 
the normal nuclear density
 $\rho_0=0.17$ fm$^{-3}$.
\label{fig:condensate}
}
\end{figure}

In Fig.~\ref{fig:condensate}, we show the calculated quark condensates
as functions of density for three types of $g_D(\rho)$ defined in
Eq.~(\ref{eq:gD}). 
In case (b), we have switched off the effect of the instanton-induced 
flavor mixing interaction and therefore the s-quark condensate has not 
changed in the SU(2) symmetric matter.  The absolute values of the 
u,d-quark condensate decreases significantly faster in case (c) than in case (a)
when the density goes up.  It means that the contribution of the 
instanton-induced interaction on the u,d-quak condensate is sizable.

\begin{figure}
\includegraphics[width=8cm]{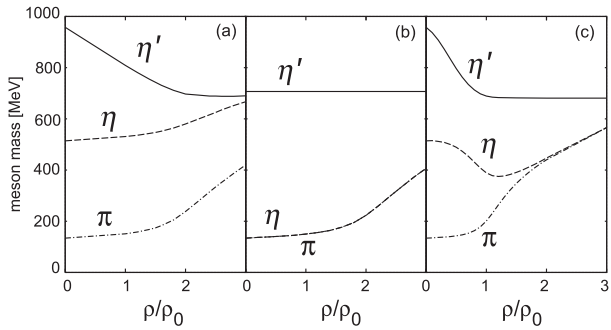}
\caption{Density dependence of the meson mass spectra. Three panels
 corresponds to the cases (a), (b) and (c) defined in Eq.~(\ref{eq:gD}),
respectively.
The nucleon density $\rho$ is
defined in Eq.~(\ref{eq:rho}) and $\rho_0$ is 
the normal nuclear density
 $\rho_0=0.17$ fm$^{-3}$.
\label{fig:meson_mass}
}
\end{figure}

In Fig.~\ref{fig:meson_mass}, we show the calculated density dependence
of the meson mass spectra for three cases defined in
Eq.~(\ref{eq:gD}). 
In the case of constant $g_D$ (a), we find that the mass of $\eta'$
decreases rapidly as 
a function of the density, while the masses of $\pi$ and $\eta$
gradually increase.
The instanton-induced interaction is repulsive for the flavor singlet
$\bar qq$ channel and the effective coupling strength 
is $g_D (\la \bar uu \ra + \la \bar dd \ra + \la \bar ss \ra)/3$.
Since the absolute values of the quark condensates decrease as the
density increase, the effective repulsive interaction in the flavor 
singlet $\bar qq$ channel becomes small as the density increases.
That is the reason why the $\eta'$ mass decrease.
In Fig.~\ref{fig:meson_mass}(b), we find that $\pi$ and $\eta$ are
degenerate completely and their masses increase gradually as density,
and the mass of $\eta'$ has no density dependence without the $U_A(1)$
anomaly effects. Without the $U_A(1)$ anomaly effects, the $\eta'$ is 
the ideal mixing state, i.e., the pure $\bar ss$, and therefore 
no density dependence in the SU(2) symmetric matter with 
$\mu_s = 0$.
It should be mentioned that the masses of $\eta$ and $\eta'$ 
at $\rho = 0$ are not reproduced in this case.
As shown in Fig.~\ref{fig:meson_mass}(c), if the $g_D$
has density dependence as case (c) in Eq.~(\ref{eq:gD}), 
we can expect mass reduction of both $\eta$ and $\eta'$ mesons 
up to 
around $\rho=\rho_0$.

It is interesting to compare our results in Fig.~\ref{fig:meson_mass}
with those obtained in Ref.~\cite{PLB634}, where the Quark-Meson
Coupling model (QMC) was used to obtain the mass-shift for given mixing
angles. The results in Table 1 of Ref.~\cite{PLB634} show that the both
masses of $\eta$ and $\eta'$ decrease in medium at normal nuclear
density (defined as $\rho_0=0.15$ fm$^{-3}$ in Ref.~\cite{PLB634}) for
all cases considered there. These results are not consistent to our
results of case (a) where the $\eta$ mass slightly increases at the
normal nuclear density. Our results of case (c) are qualitatively
consistent to those in Ref.~\cite{PLB634} and predict the mass reduction
of $\eta$ and $\eta'$ mesons in medium at normal nuclear density,
however, the sizes of the mass shift are significantly different in
these results. The origin of the discrepancies in theoretical
calculations are not clear, however, we believe that this fact indicates
the significant importantce of experimental information.

\begin{figure}
\includegraphics[width=8cm]{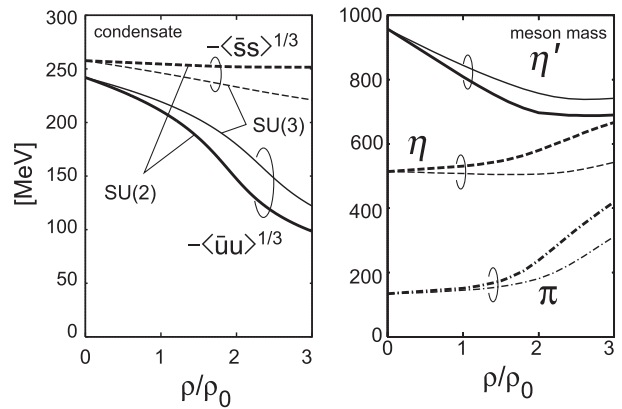}
\caption{Density dependence of the quark condensates
(left panel) and  the meson masses (right panel) are shown for the SU(2)
 symmetric matter  (thick  lines) and the  SU(3) symmetric matter (thin
 lines). The nucleon density $\rho$ is
defined in Eq.~(\ref{eq:rho}) and $\rho_0$ is 
the normal nuclear density
 $\rho_0=0.17$ fm$^{-3}$.
\label{fig:SU3}
}
\end{figure}

Here, in order to see the effects of the species of the matter to the
calculated results, we compare the results in the SU(2) symmetric matter
with those in the SU(3) symmetric matter, $\rho_u=\rho_d=\rho_s$, in
Fig.~\ref{fig:SU3}.
In Fig.~\ref{fig:SU3}, the strength of the determinant interaction is
fixed to be constant as case (a) in Eq.~(\ref{eq:gD}). 
We get same results as those obtained in Ref.~\cite{PRD71} in SU(3) case,
and find that the differences between the calculated results in SU(2)
and that of SU(3) are less than $10\%$ at normal nuclear density $\rho_0$.

\section{formation spectra of mesic nuclei}
\label{spectra}

In previous section, we discuss the possible changes of the meson
masses due to the medium effects. If such mass changes occur in the
nuclear medium, we can translate these effects into the potential forms.
We estimate the real parts of the $\eta$- and $\eta'$-nucleus
optical potentials based on the calculated their masses at finite
density by the NJL model.
The optical potential $U(r)$ 
can be written as,
\begin{equation}
U(r)=V(r)+iW(r),
\end{equation}
where $V$ and $W$ indicate the real and the imaginary parts of the
optical potential, respectively.
The mass term in the
Klein-Gordon equation for the $\eta$ and $\eta'$ mesons can be
rewritten at finite density as;
\begin{eqnarray}
m_0^2 \rightarrow m^2(\rho)&=&(m_0+\Delta m(\rho))^2 \nonumber\\
&\sim&m_0^2+2m_0\Delta m(\rho),
\label{eq:m0}
\end{eqnarray}
where $m_0$ is the calculated meson mass in vacuum
and $m(\rho)$ the meson mass at density $\rho$.
The mass shift $\Delta
m(\rho)$ is defined as $\Delta m(\rho)=m(\rho)-m_0$
and is $\Delta m(0)=0$.
Thus, we can interpret the $\Delta
m(\rho)$ as the strength of the real part of the optical potential, and
in this paper we write $V(r)$ as
\begin{equation}
V(r) = \Delta m(\rho_0)\frac{\rho(r)}{\rho_0},
\label{eq:Vr}
\end{equation}
using the meson mass shifts at normal
nuclear density $\rho_0$.
Here $\rho(r)$ is the nuclear density distribution, which is assumed to
be of an empirical Woods-Saxon form as 
\begin{equation}
\rho(r)=\frac{\rho_0}{1+\exp\left(\frac{r-R}{a}\right)},
\end{equation}
where $R=1.18A^{\frac{1}{3}}-0.48$ fm and $a=0.5$ fm with the nuclear
mass number $A$.

\begin{figure}
\includegraphics[width=8cm]{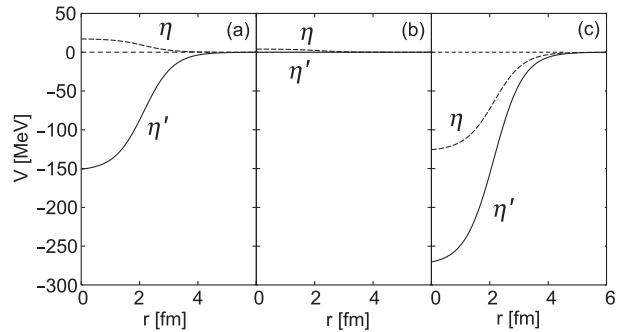}
\caption{Real parts of the $\eta$- and $\eta'$-$^{11}$B optical potentials determined by
 the meson mass shifts of NJL model in SU(2) symmetric matter as
 defined in Eq.~(\ref{eq:Vr}). Three panels 
 correspond to the cases (a), (b) and (c) defined in Eq.~(\ref{eq:gD}),
respectively.
\label{fig:ReV}
}
\end{figure}

From the calculated results shown in Fig.~\ref{fig:meson_mass}, we can
deduce the real part of the $\eta$- and $\eta'$-nucleus optical
potential as defined in Eq.~(\ref{eq:Vr}). We show the real parts of the
optical potentials in
Fig.~\ref{fig:ReV} for all these cases defined in Eq.~(\ref{eq:gD}).
We find that the sign and strength of the real part of the optical
potential strongly depends on the anomaly effects at finite $\rho$. The
potential of $\eta'$ is attractive for cases (a) and (c), and its
strength depends on the density dependence of $g_D(\rho)$. As we can see
in Fig.~\ref{fig:ReV}(b), no real potential is expected for $\eta'$
meson for $g_D=0$ case. As for $\eta$-nucleus optical potential, even
the sign of the real potential is changed by the density dependence of
$g_D(\rho)$ as shown in Figs.~\ref{fig:ReV}(a) and (c). For no anomaly
case (b), $\eta$ meson degenerates with pion and the real potential is
slightly repulsive. Thus, we confirm that we can expect to obtain the
information on $U_A(1)$ anomaly by knowing the $\eta$ and $\eta'$ meson
properties at finite $\rho$.

The strength of the imaginary potential of meson and nucleus is
estimated separately since it is extremely difficult to evaluate it using
NJL model.
For the $\eta$ meson, we estimate the strength of the imaginary
potential based on the results obtained in Refs.~\cite{our_eta} assuming
the $N^*(1535)$ dominance in the $\eta N$ channel, and take as;
\begin{equation}
W(r)=-40\frac{\rho(r)}{\rho_0} {\rm [MeV]}.
\end{equation}

As for the $\eta'$ nucleus optical potential, we can estimate $W(r)$
from analysis of 
$\gamma
p\rightarrow \eta'p$ data~\cite{nucl-th0303044}.
Since they included only $N^*(1535)$ as a baryon resonance in the
analysis of the $\eta'$ formation reaction and determined
$\eta'NN^*(1535)$ coupling strength,
we can easily
calculate the $\eta'$ 
self-energy in the medium in analogy with the $\Delta$-hole model for
the $\pi$-nucleus system as,
\begin{eqnarray}
U_{\eta'} &\sim& \frac{g^2}{2m_{\eta'}}\frac{\rho}{m_{\eta'}+M_N-M_{N^*}+i\Gamma_{N^*}/2}\nonumber\\
&=&
(+77-8i)\frac{\rho}{\rho_0}\ {\rm [MeV]}.
\label{eq:2m0U}
\end{eqnarray}
We consider the reasonable running range is
$-5$ MeV -- $-20$ MeV for the strength of the imaginary part of the
$\eta'$ nucleus optical potential
based on this evaluation in Eq.~(\ref{eq:2m0U})
as in Ref.~\cite{PRL94}.
The $\eta'$-nucleus bound states were calculated theoretically before in
Ref.~\cite{NPA670}
where the widths of the $\eta'$-mesic nuclear
states were not evaluated.

We should mention here that the evaluation in Eq.~(\ref{eq:2m0U})
provides the repulsive real part, which is opposite to the evaluation
from the $\eta'$ mass shift, since we consider only $N^*(1535)$
resonance with the mass $M_{N^*}<m_{\eta'}+M_N$ as an intermediate
states. 
By the ($\gamma$,p)
experiments proposed in this paper, we can
expect to distinguish these potentials and to determine the sign
and strength of the $\eta'$-nucleus optical potential.

\begin{figure}
\includegraphics[width=8cm]{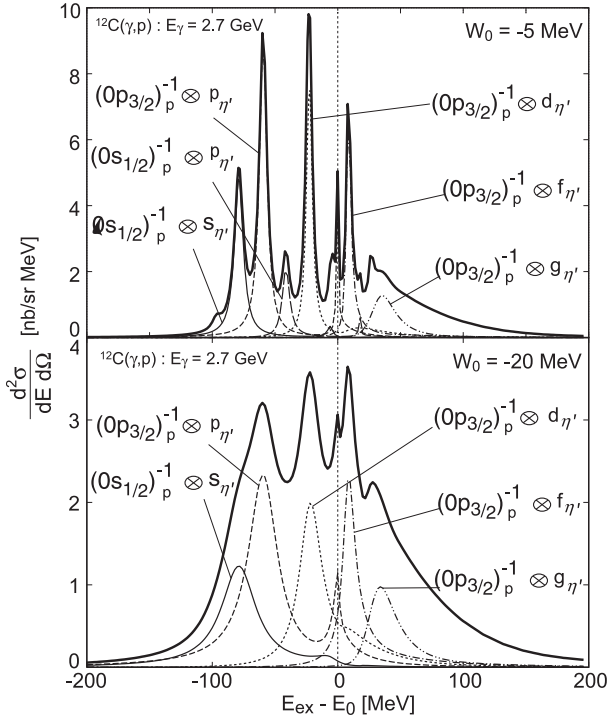}
\caption{
The calculated spectra of the $^{12}$C($\gamma$,p)$^{11}$B$\otimes\eta'$
 reaction at $E_\gamma=2.7$ GeV are shown as functions of the excited
 energy $E_{ex}$ defined in the text. $E_0$ is the $\eta'$ production
 threshold energy. The real part of the $\eta'$-nucleus optical
 potential is evaluated
for case (a) in Eq.~(\ref{eq:gD}). The
 imaginary part of the potential is assumed to be 
 $W(r)=-5\rho(r)/\rho_0$ [MeV] (upper panel) and
 $W(r)=-20\rho(r)/\rho_0$ [MeV] (lower panel). The total spectra are
 shown by the 
 thick solid line, and the dominant contributions of subcomponents are
 shown by dotted and dashed lines as indicated in the figures.
\label{fig:eta'}
}
\end{figure}

\begin{figure}
\includegraphics[width=8cm]{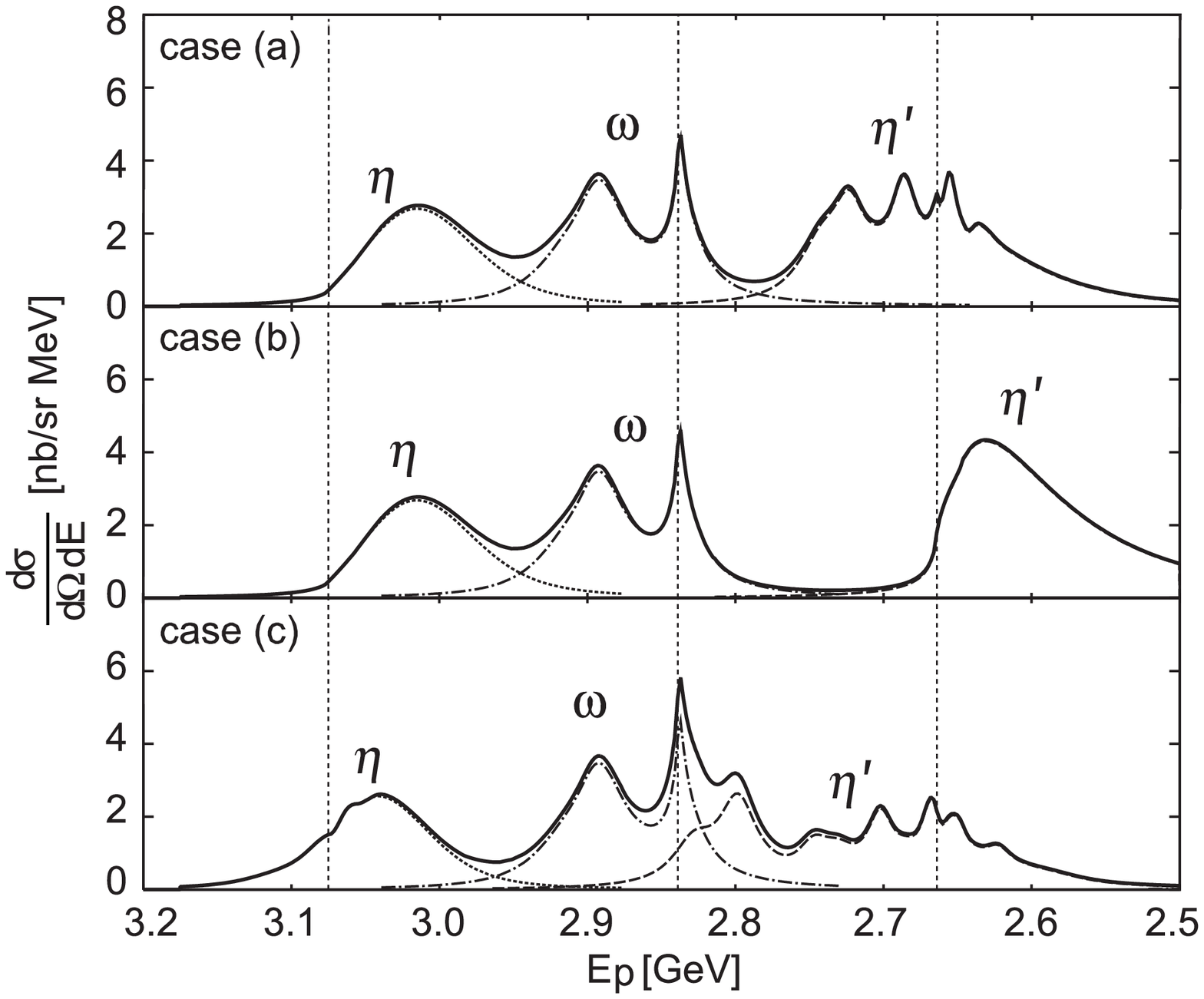}
\caption{
The calculated spectra of the
 $^{12}$C($\gamma$,p)
reactions for the $\eta$, $\omega$, $\eta'$ mesic nucleus formation at
 $E_\gamma=2.7$ GeV are shown as functions of the emitted proton
 energies in final states.
The three cases (a), (b) and (c) for $g_D$ defined in Eq.~(\ref{eq:gD})
 are considered.
The vertical dashed lines
 indicate the production thresholds of the $\eta$, $\omega$ and $\eta'$
 mesons.
The contributions from the $\eta$, $\omega$ and $\eta'$-mesic nuclei 
are shown by dotted, dash-dotted and dashed lines, respectively.
The imaginary part of the $\eta'$-nucleus optical potential is 
$W(r)=-20\rho(r)/\rho_0$ MeV.
The $\omega$-nucleus optical potential obtained in Ref.~\cite{NPA650} is
 used and is attractive.
\label{fig:with_Weise}
}
\end{figure}

\begin{figure}
\includegraphics[width=8cm]{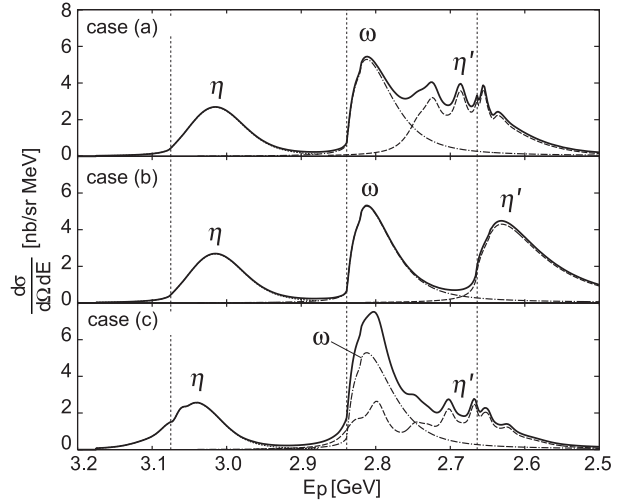}
\caption{
The calculated spectra of the
 $^{12}$C($\gamma$,p)
reactions for the $\eta$, $\omega$, $\eta'$ mesic nucleus formation at
 $E_\gamma=2.7$ GeV are shown as functions of the emitted proton
 energies in final states.
The three cases (a), (b) and (c) for $g_D$ defined in Eq.~(\ref{eq:gD})
 are considered.
The vertical dashed lines
 indicate the production thresholds of the $\eta$, $\omega$ and $\eta'$
 mesons.
The contributions from the $\eta$, $\omega$ and $\eta'$-mesic nuclei 
are shown by dotted, dash-dotted and dashed lines, respectively.
The imaginary part of the $\eta'$-nucleus optical potential is 
$W(r)=-20\rho(r)/\rho_0$ MeV.
The $\omega$-nucleus optical potential obtained in  Ref.~\cite{NPA706}
is used and is repulsive.
\label{fig:with_Lutz}
}
\end{figure}

In order to observe meson-nucleus bound states experimentally,
we consider missing mass spectroscopy~\cite{PRC44},
which was proved to be a powerful tool for the meson bound states
formation in the studies of deeply bound pionic states~\cite{PLB213etc}.
In this
spectroscopy, one observes an emitted particle in a final state, and
obtains the double differential cross section $d^2\sigma/d\Omega/dE$ as a
function of the emitted particle energy. 
As discussed in Refs.~\cite{PRL94,PLB502,PLB527,our_gamma_p}, we
think that the ($\gamma$,p) reaction with GeV photon beam is
one of
the appropriate reactions for our purpose and it can be performed in
existing facilities like SPring-8.
We consider the ($\gamma$,p) reactions for 
the $\eta$- and $\eta'$-mesic nuclei formation.

We choose the incident photon energy as $E_\gamma=2.7$ GeV, which is the
beam energy accessible at SPring-8,
and choose $^{12}$C as a target nucleus
as in Ref.~\cite{PRL94}. We use the Green
function method~\cite{NPA435} to calculate the formation cross
sections as,
\begin{equation}
\left(\frac{d^2\sigma}{d\Omega dE}
\right)_{A(\gamma,p){\eta,\eta'}\otimes(A-1)}
=\left(\frac{d\sigma}{d\Omega}
\right)^{\rm Lab}_{p(\gamma,p)\eta,\eta'}
\times
\sum_f S(E),
\label{eq:ele}
\end{equation}
where $S(E)$ is the nuclear response function and
$\left(\frac{d\sigma}{d\Omega}\right)^{\rm Lab}_{p(\gamma,p)\eta,\eta'}$ are
the elementary cross sections in the laboratory frame
for the $\eta$ and $\eta'$ meson production.
The elementary cross section for $\eta'$ production is estimated to be
$150$ nb/sr using the data of SAPHIR collaboration~\cite{PLB444} and their
analysis~\cite{PRC68}. As for the $\eta$ production process, we take the
same value for the elementary cross section as
$\left(\frac{d\sigma}{d\Omega}\right)^{\rm Lab}_{p(\gamma,p)\eta}=150$
nb/sr based on the data
at SPring-8~\cite{mura} which show the similar
production rate for both mesons in a test experiment in the preparation
stage.  
We sum up all
(proton-hole)$\otimes$(meson-particle) configurations in the final state
to get the total 
cross section in Eq.~(\ref{eq:ele}).

To calculate the response function $S(E)$, we use the Green function
$G(E;\vec{r},\vec{r'})$ defined as~\cite{NPA435},
\begin{equation}
G(E;\vec{r},\vec{r'})=\langle p^{-1}|\phi(\vec{r})
\frac{1}{E-H+i\epsilon}\phi^\dag(\vec{r'})|p^{-1}\rangle
\label{eq:Green},
\end{equation}
where $\phi^\dag$ is the meson creation operator and
$|p^{-1}\rangle$ is a proton hole state. The Hamiltonian $H$
contains the meson-nucleus optical potential $U$. We can rewrite
Eq.~(\ref{eq:Green}) in a simple expression as \cite{NPA435},
\begin{gather}
G(E;\vec{r},\vec{r'})=\sum_{l,m}
Y^*_{l,m}(\hat{r})
Y_{l,m}(\hat{r'})
G_{l}(E;r,r')\\
G_{l}(E;r,r')=-2m ku_{l}(k,r_<)v_{l}^{(+)}(k,r_>), 
\end{gather}
where $u_{l}$ and $v_{l}^{(+)}$ are the
radial part of the regular and outgoing solutions of equation of
motion, respectively. Using the Green function, the response function
can be expressed as 
\begin{multline}
S(E)=
-\frac{1}{\pi}Im\sum_{J,M,m_s}
\int d^3rd\sigma d^3r' d\sigma'\\
\times
f^\dag(\vec{r},\sigma)G(E;r,r')f(\vec{r'},\sigma).
\end{multline}
Here, we define $f(\vec{r},\sigma)$ as
\begin{equation}
f(\vec{r},\sigma)=\chi_f^*(\vec{r})\xi^*_{\frac{1}{2},m_s}(\sigma)
\left[
Y^*_{l}(\hat{r})\otimes
\psi_{j_p}(\vec{r},\sigma)
\right]_{JM}\chi_i(\vec{r}),
\end{equation}
where $\chi_i$ and $\chi_f$ denote the projectile and the
ejectile distorted waves, respectively, $\psi$ is the proton-hole
wavefunction and 
$\xi$ is the spin wavefunction introduced to count possible spin
directions of the proton in the target nucleus. 
The distorted waves $\chi_i$ and $\chi_f$ are
evaluated by using the eikonal
approximation as in Ref.~\cite{PRL94}.

In Fig.~\ref{fig:eta'}, we show the calculated spectra
of $^{12}$C($\gamma$,p) reaction for the $\eta'$ meson formation
as functions of the excited energy which are defined as,
\begin{equation}
E_{ex}=m_{\eta'}-B_{\eta'}+[S_p(j_p)-S_p(p_{3/2})],
\end{equation}
where $B_{\eta'}$ is the $\eta'$ binding energy and $S_p$ the proton
separation energy. The $\eta'$ production threshold energy $E_0$ is
indicated in the figure by the vertical dashed line.
The real part of the optical potential is evaluated
for case (a) in
Eq.~(\ref{eq:gD})
which has around $-150$ MeV depth
at nuclear center as
can be seen in Fig.~\ref{fig:ReV}(a). 
The spectra are calculated for (upper panel) $W_0=-5$ MeV and (lower
panel) $W_0=-20$ MeV cases where the imaginary potential is written as
$W(r)=W_0\rho(r)/\rho_0$.
We also show the dominant subcomponents of the
spectra in Fig.~\ref{fig:eta'}.
As we can see from these figures, we can expect to observe the peak
structures in the spectra due to the formation of the $\eta'$-mesic
nucleus
even in the case with the relatively large imaginary potential,
and we can expect to deduce the magnitude of the $\eta'$ mass
shift at finite nuclear density from the observed spectra. 
The evaluated imaginary part of the $\eta'$-nucleus potential is small
enough 
and the resonance peaks are expected to be clearly separated each
other. 
Hereafter, we show the calculated spectra of the $\eta'$-mesic nuclei
formation only for
$W_0= -20$ MeV cases.

In Figs.~\ref{fig:with_Weise} and \ref{fig:with_Lutz}, we show the
($\gamma$,p) spectra for wider energy region including 
both $\eta$ and $\eta'$ production to see the anomaly effects in the
whole ($\gamma$,p) spectra simultaneously. To provide realistic spectra,
we include the contributions due to $\omega$ meson production processes,
which exist between $\eta$ and $\eta'$ contributions. As the
$\omega$-nucleus interaction, we apply the results obtained by two
theory groups reported in Refs.~\cite{NPA650} and \cite{NPA706}.
We should mention here that the $\omega$-nucleus interaction at
threshold in Ref.~\cite{NPA650} is attractive, while that in
Ref.~\cite{NPA706} is repulsive. Hence, we use both interactions to
evaluate the contribution of $\omega$ meson. 

In Fig.~\ref{fig:with_Weise}, we show the spectra including the
contribution due to the formation of the $\omega$ mesic nuclei with the
attractive potential reported in Ref.~\cite{NPA650}.
We can see the significant mass reduction of the $\eta'$ meson in
nuclear medium in
Fig.~\ref{fig:with_Weise}(a), while in
Fig.~\ref{fig:with_Weise}(b) we only see the simple quasi-free
$\eta'$ formation spectrum without $U_A(1)$ anomaly effect.
In the case with the density dependent $g_D$ in
Fig.~\ref{fig:with_Weise}(c),  
we can see the mass reduction of both the $\eta$ and $\eta'$ mesons.
Especially, even below the $\omega$ production threshold,
we can see the finite contributions of the $\eta'$ mesic nuclei
formation to the spectrum as the consequences of the large $\eta'$
mass reduction in the case with the density dependent $g_D$ defined as
(c) in Eq.~(\ref{eq:gD}).
For the $\eta$ mesic nuclei production, it is hardly to see the
differences between the cases with (a) the constant $g_D$ and (c) the
density dependence $g_D$, due to the large momentum transfer for the
$\eta$ mesic nuclei formation with the incident photon energy
$E_\gamma=2.7$ GeV considered in this paper.
At this energy, the contributions from higher partial waves of 
quasi-free $\eta$ meson production
are so large that those from the bound $\eta$ states hardly affect the  
total spectrum.
Furthermore, we should make a comment on the $\eta$ mesic nuclei that,
in order to see the anomaly effect clearly, we have not considered any
contributions 
from resonances like $N^*(1535)$ in this paper in the estimation of the
real part of 
the $\eta$-nucleus optical potential, whose contribution is known
to be important on the $\eta$-nucleon system. The detailed discussions
of the $\eta$-mesic nuclei
with the $N^*(1535)$ resonance
are reported in 
Refs.~\cite{our_eta,our_gamma_p}.
The calculated contribution of the $\omega$ channel in
Fig.~\ref{fig:with_Weise} is essentially the same as that reported in
Ref.~\cite{PLB502}.

Similarly, in Fig.~\ref{fig:with_Lutz} we show the spectra with the
$\omega$ mesic nuclei with the 
repulsive potential predicted in Ref.~\cite{NPA706}. In this case, the
contribution of the quasi-free $\omega$ formation locates the same energy region
with the bound $\eta'$ in the cases (a) and, especially (c) in
Fig.~\ref{fig:with_Lutz}. In Fig.~\ref{fig:with_Lutz}(b), we
can only see simple three peaks corresponding the quasi-free $\eta$,
$\omega$ and $\eta'$ mesons in the spectrum.


As for
the background, which is very important to discuss
experimental feasibility, is evaluated by using the experimental data
taken by LEPS collaboration at SPring-8 recently~\cite{mura}. 
That was a test experiment in the preparation stage for the observation
of the $\omega$ mesic nuclei by the ($\gamma$,p) reaction, which used the
same kinematics proposed in this paper and observed the background
proton emission rate from Carbon target including energy region for
the $\eta'$ meson production~\cite{mura}.  We can roughly estimate the
order of magnitude of the background proton cross section as to be
$10$ -- $100$ [nb/sr MeV] in the $\eta'$ formation region.  Thus, we 
estimate the signal over noise ratio
is around $S/N \sim 1/10$. We think the absolute magnitude of the 
calculated formation cross section is reasonably large and the spectra
are expected to be observed in future experiments at SPring-8~\cite{mura}.

\section{Conclusions}
\label{conc}
We have investigated the possible observation of the effective
restoration of the $U_A(1)$ anomaly at finite density by formations of
the $\eta$- and 
$\eta'$-mesic nuclei. Especially, since the heavy $\eta'$ meson mass is
considered to be generated from the $U_A(1)$ anomaly effect, we can
expect to observe clearly the anomaly effect in medium in the $\eta'$
channel. 

In order to evaluate more quantitative magnitude of the meson mass
changes 
due to the medium effects
and to obtain deeper insights on $U_A(1)$ anomaly effects than
Ref.~\cite{PRL94},
we adopted the Nambu-Jona-Lasinio 
(NJL) model with the Kobayashi-Maskawa-'t Hooft (KMT) interaction, 
and calculated the meson masses in finite density with three different 
coupling strengths of the KMT interaction.
We obtained
significant $\eta'$ mass changes even at normal nuclear density due to
the effective restoration of the $U_A(1)$ anomaly. 
In the finite density calculations, we have substituted the SU(2) symmetric 
constituent quark matter for the symmetric nuclear matter.
Although the NJL model doesn't confine quarks, it is a
good effective model which can describe the low energy
QCD phenomena, i.e., the properties of light pseudoscalar
mesons rather well after fixing the model parameters
appropriately.  Even the NJL model has some shortcomings,
we think it is meaningful to apply the model in this exploratory level
to clarify the physics of $\eta' (958)$ mesic nuclei and $U_A(1)$
anomaly at finite density.

To investigate the experimental feasibility, we calculated 
formation cross sections of the $\eta$- and $\eta'$-mesic nuclei with
($\gamma$,p) reaction.
We found that the calculated cross section has reasonably large
magnitude, and the ($\gamma$,p) reaction with GeV photon beam, which
can be provided in existing facilities like SPring-8, is an appropriate
reaction for our purpose. We conclude that we can expect to observe the
$\eta'$ mass reduction clearly in this reaction, and to obtain new
information on the $U_A(1)$ anomaly at finite density.

The present evaluation is the first theoretical results for the
formation reaction of the $\eta$- and $\eta'$-mesic nuclei
based on the NJL model results
to know the behavior of
$U_A(1)$ 
anomaly in the medium. We believe that the present theoretical results
is much important to stimulate both theoretical and experimental
activities to study the $U_A(1)$ anomaly at finite density and to obtain
the deeper 
insights of QCD symmetry breaking pattern and the meson mass spectrum.

\section*{Acknowledgement}
We would like to thank 
D.~Jido and M.~Oka for
useful comments.
We also thank to N.~Muramatsu for valuable discussions on the latest
data of the ($\gamma$,p) reactions at SPring-8.
H.~N. and S.~H. also thank to T. Hatsuda for valuable discussions at the
early stage of this work.
H.~N. thanks to A.~Hosaka for useful comments and discussions.
One of the authors (H.~N.) is supported by Research Fellowships of the
Japan Society for the Promotion of Science (JSPS) for Young Scientists.
This work is partly supported by Grants-in-Aid
for scientific research of MonbuKagakushou and Japan Society for the
Promotion of Science (No.~16540254 (S.~H.) and No.~18$\cdot$8661 (H.~N.)).


\end{document}